\documentclass[useAMS,usenatbib]{mn2e}
\usepackage{amsmath}
\usepackage{amssymb}
\usepackage{graphicx}
\usepackage{hyperref}

\newcommand{\be}{\begin{equation}}
\newcommand{\ee}{\end{equation}}
\newcommand{\beq}{\begin{eqnarray}}
\newcommand{\eeq}{\end{eqnarray}}
\newcommand{\n}{\mathrm{n}}
\newcommand{\cc}{\mathrm{p}}

\newcommand{\x}{\mathrm{x}}
\newcommand{\y}{\mathrm{y}}
\newcommand{\vv}{\mathrm{v}}

\title[The effect of SF hydro on glitches ]{The effect of superfluid hydrodynamics on pulsar glitch sizes and waiting times}
\author[B.Haskell]{B.~Haskell\\
School of Physics, The University of Melbourne, Parkville, VIC 3010, Australia\\
Nicolaus Copernicus Astronomical Center, Polish Academy of Sciences, Bartycka 18, 00-716, Warszawa, Poland}

\begin{document}

\pagerange{\pageref{firstpage}--\pageref{lastpage}} \pubyear{2014}

\maketitle

\label{firstpage}

\begin{abstract}

Pulsar glitches, sudden jumps in frequency observed in many radio pulsars, may be the macroscopic manifestation of superfluid vortex avalanches on the microscopic scale. Small scale quantum mechanical simulations of vortex motion in a decelerating container have shown that such events are possible and predict power-law distributions for the size of the events, and exponential distributions for the waiting time. Despite a paucity of data, this prediction is consistent with the size and waiting time distributions of most glitching pulsars. Nevertheless a few object appear to glitch quasi-periodically, and exhibit many large glitches, while a recent study of the Crab pulsar has suggested a cut-off deviations from a power-law distribution for smaller glitches \citep{CrabSize}. In this paper we incorporate the results of quantum mechanical simulations in a macroscopic scale superfluid hydrodynamics  simulation. We show that the effect of vortex coupling to the neutron and proton fluids in the neutron star naturally leads to deviations from power-law distributions for sizes and from exponential distributions for waiting times. In particular we predict a cut-off in the size distribution for small glitches. 

\end{abstract}


\section{Introduction}
 
The rotation rate of radio pulsars, rotating magnetised Neutron Stars (NSs), is exquisitely stable, in some cases rivalling the stability of atomic clocks. However some pulsars undergo sudden jumps in frequency, known as `glitches', that are instantaneous to the accuracy of the data. The origin of these events is still debated forty years after their first discovery, but it is generally thought that they are due to a large scale superfluid  component in the NS interior that is only weakly coupled to the `normal' component of the star which is tracked by the radio signal. On theoretical grounds neutrons are expected to be superfluid, as a mature NS is likely to be cold enough for most regions in the interior to be below the superfluid transition temperature \citep{BPPsuper}. Superfluidity has a strong impact on the dynamics of the system, as a superfluid rotates by forming an array of quantised vortices, which mediate a dissipative interaction between the superfluid and the normal fluid; the mutual friction. Vortices, however, can also be strongly attracted, or `pinned', to ions in the crust \citep{AlparPin1} or superconducting flux tubes in the core of the star \citep{Link03}. In this case the superfluid cannot expel vorticity and cannot spin down together with the normal component, thus lagging behind and storing angular momentum. The sudden re-coupling of the two components leads to an exchange of angular momentum and a glitch, as was suggested early on by \citet{AI}.

Despite the success of this paradigm in explaining many qualitative features of pulsar glitches, and the bulk of work that has been devoted to studying the response of the star to a glitch (see \citet{HMReview} for a review), it is still unclear what triggers such an event. Several mechanisms have been suggested, including starquakes \citep{Rud69}, superfluid instabilities \citep{prix03} and vortex avalanches \citep{Alpartrigger, Melatos08}. This last mechanism is based on the idea that vortices in a NS may form a Self Organised Critical (SOC) system, in which global stresses, due to classical drag forces acting on pinned vortices, are relieved locally via discrete avalanches, so that the system self-regulates and is always close to the critical threshold for unpinning \citep{Lila08, Lila12}. As a consequence of self organised criticality, the distribution of glitch sizes in a pulsar is expected to be a power law and the distribution of waiting times an exponential. This is approximately true for many glitching pulsars \citep{Melatos08}, although the small number of events does not allow for strong statistical conclusions. There are, however, at least three pulsars, the best known example being the Vela pulsar, which appear to mostly have glitches of a typical size which occur quasi-periodically \citep{Melatos08, George}. This behaviour is more reminiscent of the so-called `snowplow' mechanism \citep{pierre}, in which the global reservoir of angular momentum is depleted periodically once vortices can no longer be held in place by the pinning force. The periodicity and size of Vela glitches can be well reproduced in this framework, as the system has a natural length-scale and time-scale set by the height and position of the maximum of the microphysical pinning force. Additionally, a recent analysis of glitches in the Crab pulsar suggests that there may be a deviation from a power law distribution for the sizes, and that glitches may have a substantial minimum size \citep{CrabSize}.

In this paper we investigate how local, small scale, distributions of unpinning events are affected by large scale hydrodynamics and coupling between the fluids. In particular we will show that even if unpinning events  are distributed as a power-laws on a microscopic level, the large scale glitch distribution can be substantially different, and exhibit a cutoff for small glitch sizes.

\section{Methods}

We model the NS as a two-fluid system of superfluid neutrons and a charge neutral component consisting of the crust and electromagnetically bound protons and electrons. We thus do not consider vortex motion directly, but rather average over many vortices to consider the macroscopic motion of two dynamical degrees of freedom. Following \citet{AC06} we can write conservation laws for each species $\x$:
\be
\partial_t \rho_\x+\nabla_i (\rho_x v_\x^i)=0,
\ee
where $\rho_x$ is the density of constituent $\x$ (with $\x=\n$ for superfluid neutrons and $\x=\cc$ for the locked proton,electron and crust fluid). The Euler equations are:
\be
(\partial_t+v_j^\x\nabla_j)(v_i^\x+\varepsilon_\x w_i^{\y\x})+\nabla_i(\tilde{\mu}_\x+\Phi)+\varepsilon_\x w_{\y\x}^j\nabla_j v_j^\x=f_i^\x/\rho_\x,
\ee 
where we assume sums over repeated spacial indices, $w_i^{\y\x}=v_i^\y-v_i^\x$, $\varepsilon_\x$ is the entrainment parameter, $\Phi$ is the gravitational potential and $\tilde{\mu}_\x=\mu_\x/m_\x$ is the chemical potential per unit mass (and we will assume $m_\cc=m_\n$). Finally $f_i^\x$ is the mutual friction force, which for straight vortices takes the form:
\be
f_i^\x=\gamma \kappa\n_\vv \rho_\n \mathcal{B}^{'}\epsilon_{ijk}\hat{\Omega}_\n^j w_{\x\y}^k+\gamma\kappa n_\vv\rho_\n\mathcal{B}\epsilon_{ijk}\hat{\Omega}_\n^j\epsilon^{klm}\hat{\Omega}^\n_l w_m^{\x\y},
\ee
with $\Omega^j$ the angular frequency of the neutron fluid and $\kappa=h/2 m_\n$ the quantum of circulation. For the vortex number per unit area $n_\vv$ one has the relation:
\be
\kappa n_\vv=2\left[\Omega_\n+\varepsilon_\n(\Omega_\cc-\Omega_\n)\right]+\tilde{r}\partial_{\tilde{r}}\left[\Omega_\n+\varepsilon_\n (\Omega_\cc-\Omega_\n)\right]
\ee
where $\tilde{r}$ is the cylindrical radius and $\gamma<1$ represents the fraction of vortices which are free \citep{JM06}, with the remaining vortices pinned. The strength of the mutual friction is parametrised by the dimensionless constants $\mathcal{B}$ and $\mathcal{B}^{'}$.

Following \citet{Haskell12} we consider the two components to be rotating around the same axis defined by $\hat{\Omega}_\cc$, and assume that the proton fluid is rigidly rotating on the timescales of interest. This assumption is justified on longer post-glitch timescales, on which the crust can be considered as a solid and the magnetic field of the star will lock it to the protons in the core. On shorter timescales, however, several modes of oscillation of the fluids could be present \citep{Sidery10, vE14}, and are neglected in the current treatment. The equations of motion, averaged over vortex length (assuming straight vortices aligned with the $z$ axis), take the form:
\beq
\dot{\Omega}_\cc&=&-\dot{\Omega}_O+\int \frac{\tilde{Q}(\tilde{r})}{I_\cc}\frac{\tilde{r}^2}{1-\varepsilon_\n-\varepsilon_\cc} dV\label{evo1}\\
\dot{\Omega}_\n&=&- \frac{\tilde{Q}(\tilde{r})}{\rho_\n}\frac{1}{1-\varepsilon_\n-\varepsilon_\cc} dV - g (\varepsilon_\n) \dot{\Omega}_O+\tilde{F}_p\label{evo2}
\eeq
where we have defined $\tilde{Q}(\tilde{r})=\rho_\n\gamma\kappa n_\vv\mathcal{B}(\Omega_\cc-\Omega_\n)$ and $\dot{\Omega}_O$ is the contribution to the spin evolution from the external spin down torque acting on the star. In the following we take $g(\varepsilon_\n)=0$. This is not accurate in the crust, as entrainment will be strong and reduce the amount of angular momentum available for a glitch \citep{Chamel1, Chamel2, crust not}. For our purposes this is, however, simply a rescaling of the allowed size of glitches, and as we are interested in distributions and not in fitting absolute sizes, we will neglect this term for computational ease. We have also included the contribution due to pinning, $\tilde{F}_p$, which we define as:
\beq
\tilde{F}_p&=& \frac{\tilde{Q}(\tilde{r})}{\rho_\n}\frac{1}{1-\varepsilon_\n-\varepsilon_\cc} dV\;\;\;\mbox{for}\;\;\; \Omega_\n-\Omega_\cc<\Delta_c\\
\tilde{F}_p&=&0\;\;\;\mbox{for}\;\;\; \Omega_\n-\Omega_\cc\geq\Delta_c
\eeq
where $\Delta_c$ is the critical lag for unpinning, and must be determined from microphysical calculations of pinning forces. We approximate the realistic results of \citet{Sevesopin} by taking a Gaussian profile of the form:
\beq
\Delta_c(\tilde{r})\!\!\!\!\!&=&\!\!\!\!\!\Delta_M \exp{\left(-\frac{(\tilde{r}-r_m)^2}{2\sigma^2}\right)}+\Delta_{min}\label{delta1} \;\;\;\mbox{for}\;\;R_c<\tilde{r}\\
\Delta_c(\tilde{r})\!\!\!\!\!&=&\!\!\!\!\!\frac{\Delta_{min}-\tau\dot{\Omega}_O}{R_c-R_i}( \tilde{r}-R_i)+\tau\dot{\Omega}_O\;\;\mbox{for}\;\;R_i<\tilde{r}\leq R_c
\eeq
where $r_m$ is the location of the maximum, which we take to be at $\rho=7\times 10^{13}$ g/cm$^3$, and we take as typical values $\Delta_M=10^{-2}$, $\Delta_{min}=2\times 10^{-4}$ and $\sigma=10^{-2} (r_m-R_c)$. We model the equation of state as an $n=1$ polytrope. The timescale $\tau$ is the minimal coupling timescale that we resolve in our numerical formulation, which we fix as $\tau=60 s$, to approximate the observational upper limits on the rise time of a glitch  \citep{Dodson07}. The external boundary of our simulations $R_{nd}$ is the neutron drip radius and $R_c$ is the the radius of the crust-core interface (taken at $\rho=1.6\times 10^{14}$ g/cm$^3$). $R_i$ is the internal boundary, which we take to be the radius at which $\tau_{MF}=\tau$, with $\tau_{MF}=1/(2\Omega_\n<\mathcal{B}>)$, with $<\mathcal{B}>$ the standard mutual friction coefficient for electron-vortex scattering from \citet{TrevMF}, averaged over the vortex length \citep{Haskell12}. The moment of inertia $I_\cc$ in equations (\ref{evo1})-(\ref{evo2}) is thus the combined moment of inertia of the outer crust, protons in the computational domain, and all components for $\tilde{r}\leq R_i$.

Finally, we fix the initial value of $\gamma\mathcal{B}$ over the computational domain to $\gamma\mathcal{B}={\dot{\Omega}_O}/{2\Omega_\n\Delta_c}$, which ensures that the neutrons are spinning down together with the proton fluid, with a fixed lag close to the critical lag. This not only makes certain that the system is sub-critical from the start, but also allows us to circumvent the substantial uncertainties in microphysical estimates of $\mathcal{B}$ in the crust.

\subsection{Random unpinning and avalanches}

Evolving the system of equations in (\ref{evo1})-(\ref{evo2}), as described in the previous section, will not lead to glitches. An unpinning trigger has to be added to initiate a glitch. \citet{Haskell12} followed the so-called `snowplow' model \citep{pierre} and assumed that a vortex sheet forms close to the maximum of the critical lag $\Delta_c$. Once said lag is exceeded vortices are free to move out, leading to an increase in $\mathcal{B}$ (assumed to be due to Kelvin waves being excited by rapid vortex motion) and a glitch. This mechanism predicts well the sizes and waiting-times of the Vela and other pulsars that exhibit mainly giant glitches, however it cannot explain why the sizes of glitches in many other pulsars span several decades and are consistent with a  power-law distribution \citep{Melatos08}. On the other hand quantum mechanical Gross-Pitaevskii (GP) simulations of pinned vortices in a decelerating trap show that vortex avalanches can lead to glitches and naturally give rise to power-law distributions for their sizes \citep{Lila11}. In this paper we take a first step towards reconciling these two approaches by taking the large scale two-fluid hydrodynamical NS model described in the previous section as a background over which to evolve small scale fluctuations in the number of pinned vortices, as predicted by GP simulations.

In practice we will follow the approach of \citet{HA14} and assume that vortex avalanches can randomly unpin vortices, increasing the unpinned fraction from $\gamma$ to $\alpha\gamma$ in a region $R_{nd}-R_{av}<r<R_{nd}$, where both $\alpha$ and $R_{av}$ are drawn from a power law distribution of the form
\be
p(x)=\frac{(k-1)x^k}{x_M^{k+1}-x_m^{k+1}}\label{power}
\ee
where $x=R_{av},\alpha$; the power-law index is $k$, and $x_m=1, x_M=10^6$. We also assume that waiting times $t$ between events are exponentially distributed:
\be
p(t)=\frac{1}{t_w}\exp{(-t/t_w)}
\ee
with $t_w$ the mean waiting time between unpinning events, which we stress is not necessarily the mean waiting time between observed glitches.

We thus draw a value $\alpha>1$ for the region $R_{nd}-R_{av}<r<R_{nd}$, and the increased coupling leads to a `glitch', although for small values of $\alpha$, the event is slow enough and weak enough that it will not appear as a sudden  jump in frequency, but as a gradual increase of the spin-down rate.

Following the results of \citet{hopping}  we assume that vortices can repin if the lag $\Omega_\n-\Omega_\cc\leq 0.98 \Delta_C$. In practice once the lag has been reduced below this limit, we set $\gamma=0$, until the lag increases again to $\Omega_\n-\Omega_\cc = \Delta_C$, when we set once again $\gamma=1$. 

Let us note that before running the hydrodynamical simulations we have verified that simply drawing $R_{av}$ from the distribution in (\ref{power}) and transferring the angular momentum of the superfluid in the region $R_{nd}-R_{av}<r<R_{nd}$ to the crust leads to a power-law distribution for the glitches for which we recover the original microphysical index $k$, and the smallest glitch size we can resolve is $\Delta\Omega_\cc/\Omega_\cc=8.7\times 10^{-14}$.

\section{Results}

\begin{figure}
\centerline{\includegraphics[scale=0.5]{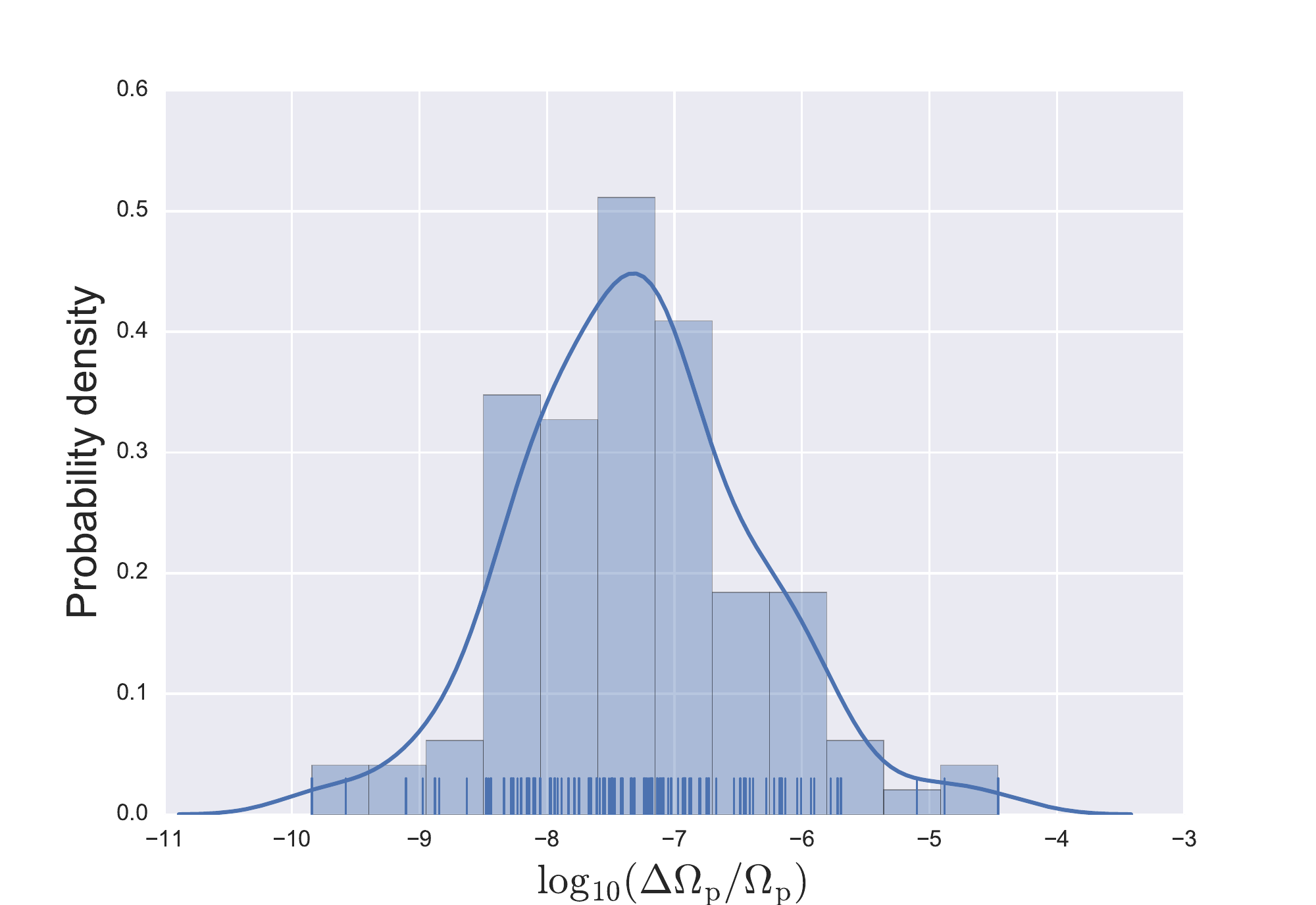}}
\caption{Probability distribution function for glitch sizes $\Delta\Omega_\mathrm{p}/\Omega_\mathrm{p}$ for a model with microscopic power-law index $k=-1.05$ and waiting time $t_w=0.1$ days. The probability distribution function is estimated both with a histogram and a gaussian kernel density estimator. The distribution deviates significantly from a power law at lower sizes, with a notable absence of small glitches. This is due to the fact that very small events do not appear as sudden jumps, i.e. glitches, but rather as gradual changes in the spin-down rate (i.e. timing noise) and are not flagged by our glitch-finding algorithm. }\label{fig1}
\end{figure}
\begin{figure}
\centerline{\includegraphics[scale=0.5]{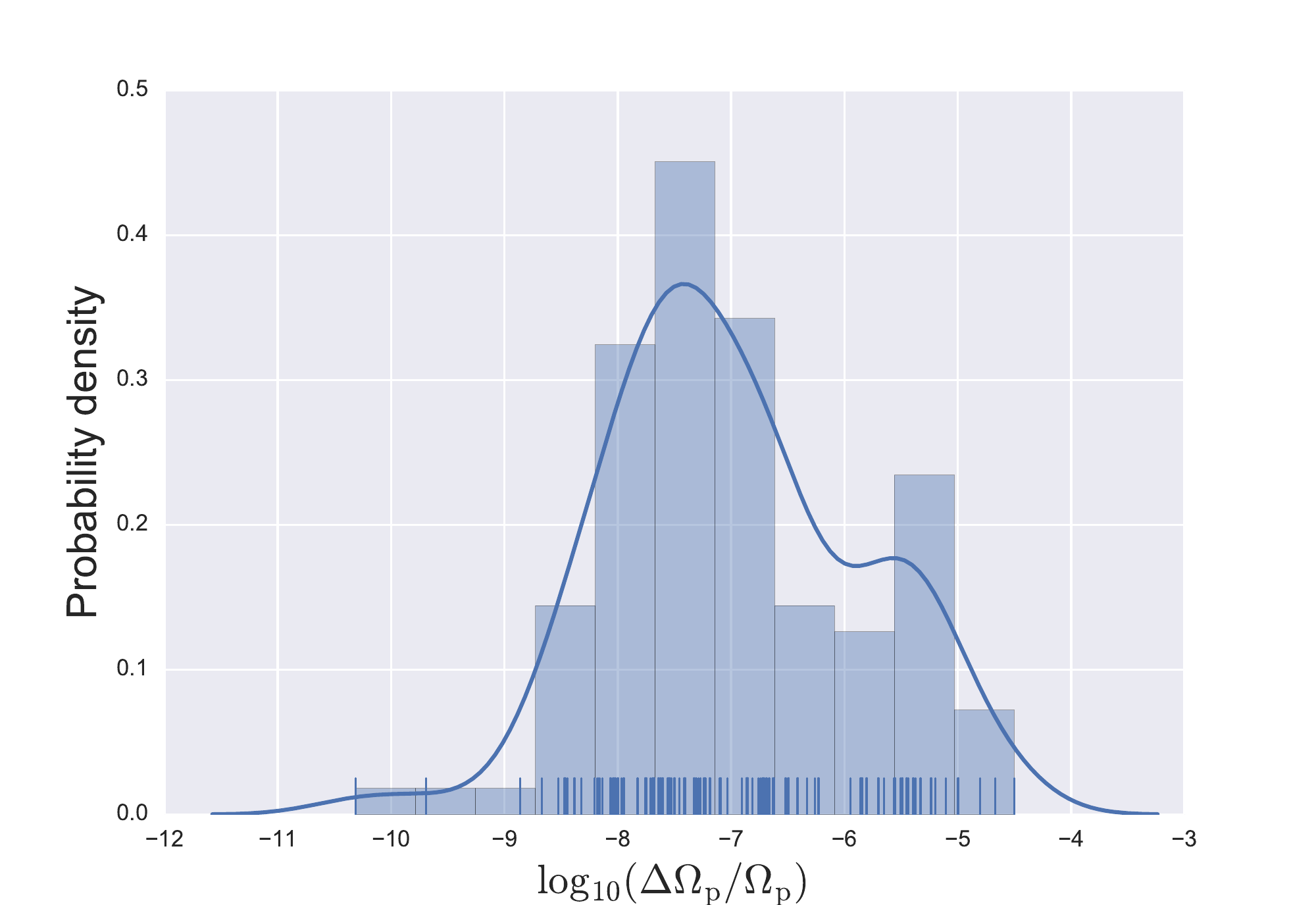}}
\caption{Probability distribution function for glitch sizes $\Delta\Omega_\mathrm{p}/\Omega_\mathrm{p}$ for a model with microscopic power-law index $k=-1.5$ and waiting time $t_w=0.1$ days. The probability distribution function is estimated both with a histogram and a gaussian kernel density estimator. As in the previous case with $k=-1.05$ the distribution deviates significantly from a power law at the lower end, but there is also an excess of large glitches. }\label{fig2}
\end{figure}

\begin{figure}
\centerline{\includegraphics[scale=0.45]{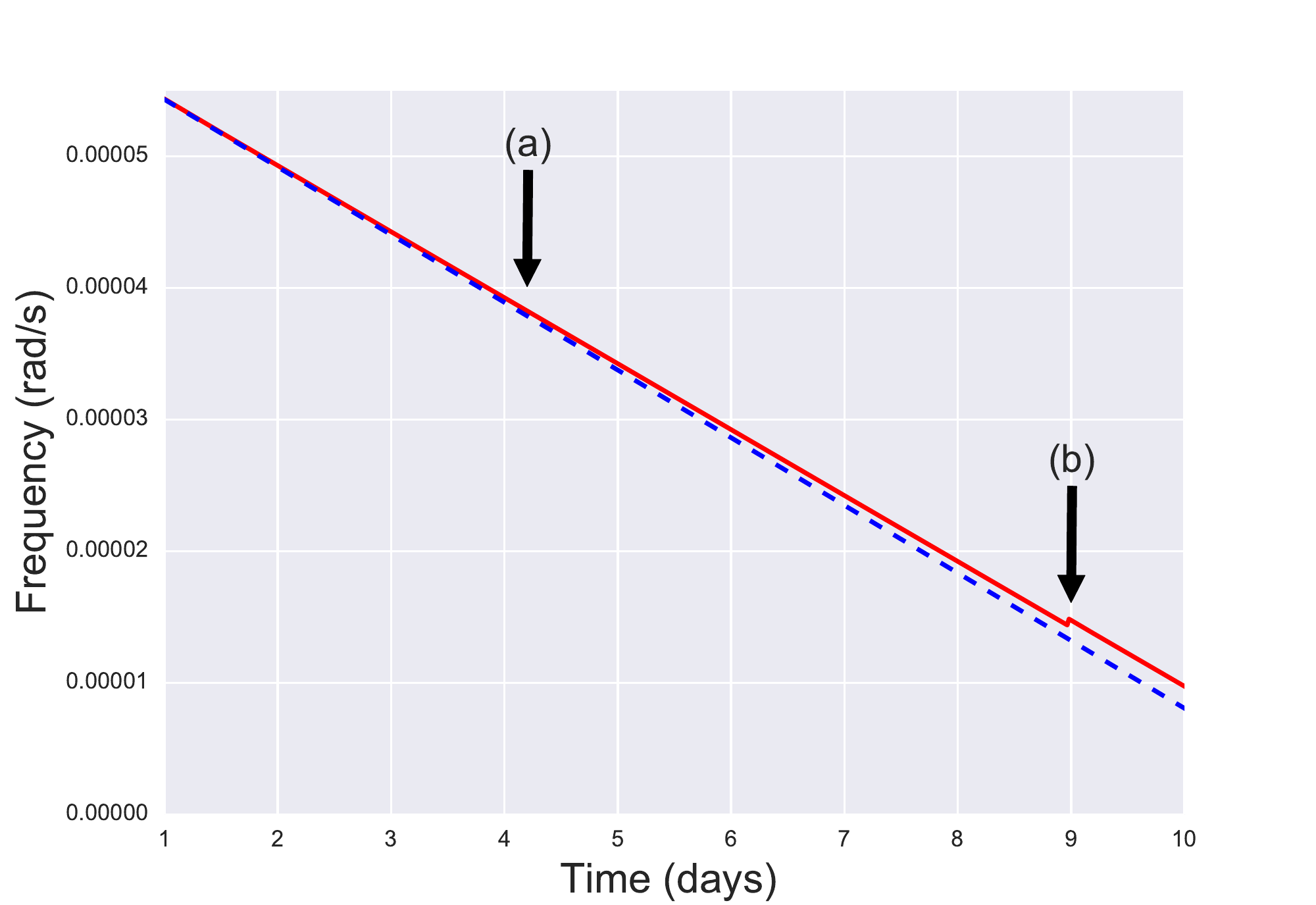}}
\caption{Example of a spin down curve obtained from a simulation with $k=-1.05$, $t_w=1$ day (thick red line), compared to a linear fit to the spin down from previous data (dashed blue line). Two events are notable, event (a) which corresponds to a slow decrease of the spin down rate, and is not detected by the glitch finder, and event (b), which is a standard glitch.}\label{fig3}
\end{figure}

We have run a number of simulations, each comprising approximately 100 glitches, with varying $\Delta_C$, mass $M$, mean waiting time $t_w$ and power-law index $k$ and analysed the results. After running the simulations we run a simple glitch finding algorithm, that identifies all events in which the spin frequency of the 'crust' $\Omega_\cc$ rises, i.e. $\dot{\Omega}_\cc>0$, and measures the maximum size of the event from the start of the rise to the maximum of the frequency. The main conclusion is that, independently of the choice of parameters,  the macroscopic glitch size and waiting time distributions that we extract from the simulation differ significantly from our microphysical inputs. For all cases size distributions deviate from power-law distributions at the lower end, with a strong drop off in the number of glitches for smaller sizes, as can be seen from the examples in figures \ref{fig1} and \ref{fig2}. This is due to the fact that for small values of $\alpha$, the effective mutual friction parameter $\gamma\mathcal{B}$ is small, leading to a slow, more gradual event that also exchanges a small amount of angular momentum, due to the low value of $R_{av}$. This event is thus a gradual change in the spin-down rate, more similar to timing-noise, and is not identified as a glitch by our glitch finding algorithm. An example of such an event is shown in figure \ref{fig3}. 

This is consistent with observations of glitches in the Crab pulsar, which reveal deviations of the size distribution from a power-law with a lack of observed small glitches\citep{CrabSize}. Our results show that the effect of superfluid hydrodynamics are sizeable and lead to deviations from power law distributions, even though on a microphysical scale vortex unpinning events may still be power-law distributed. 

\begin{figure*}
\centerline{\includegraphics[scale=0.6]{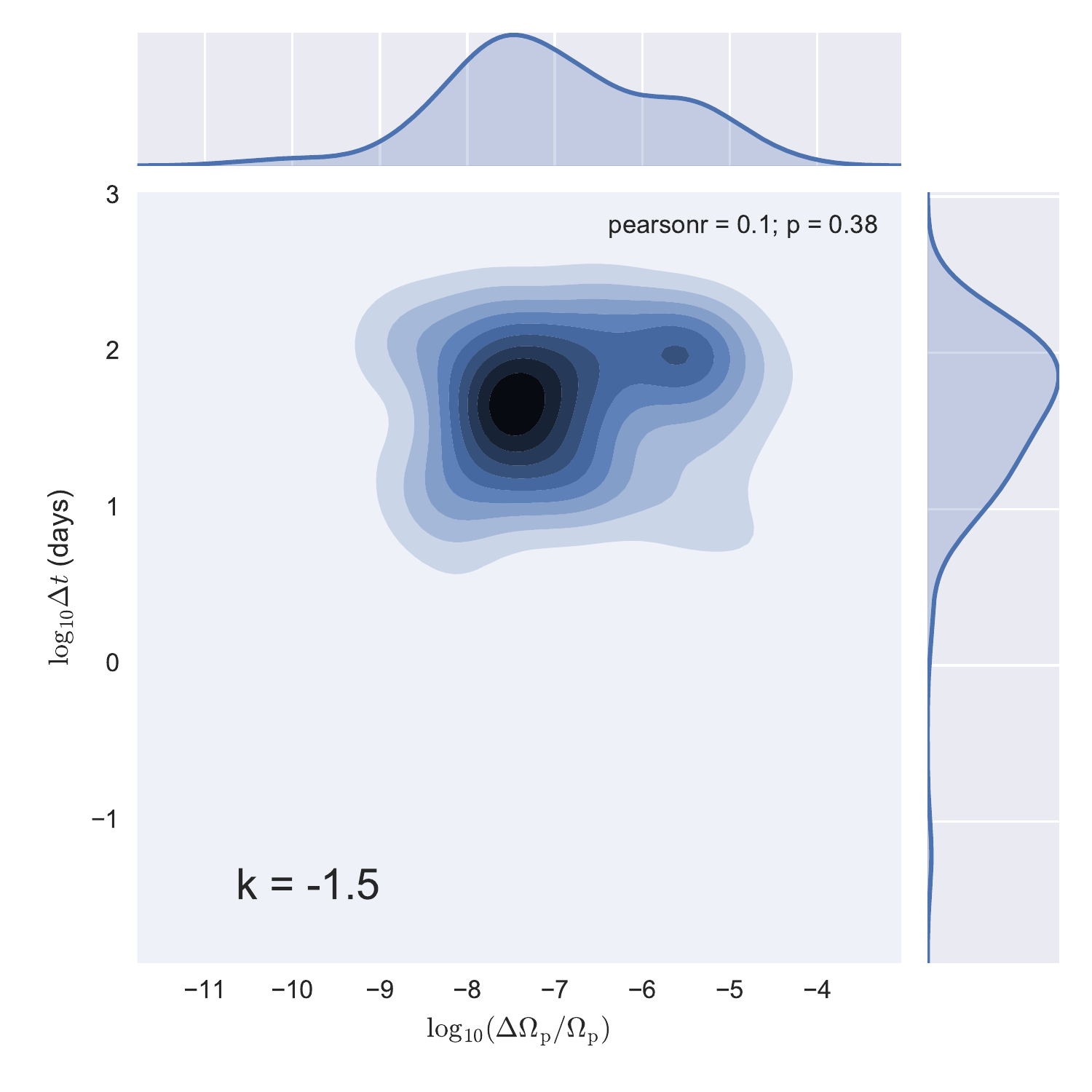}\includegraphics[scale=0.6]{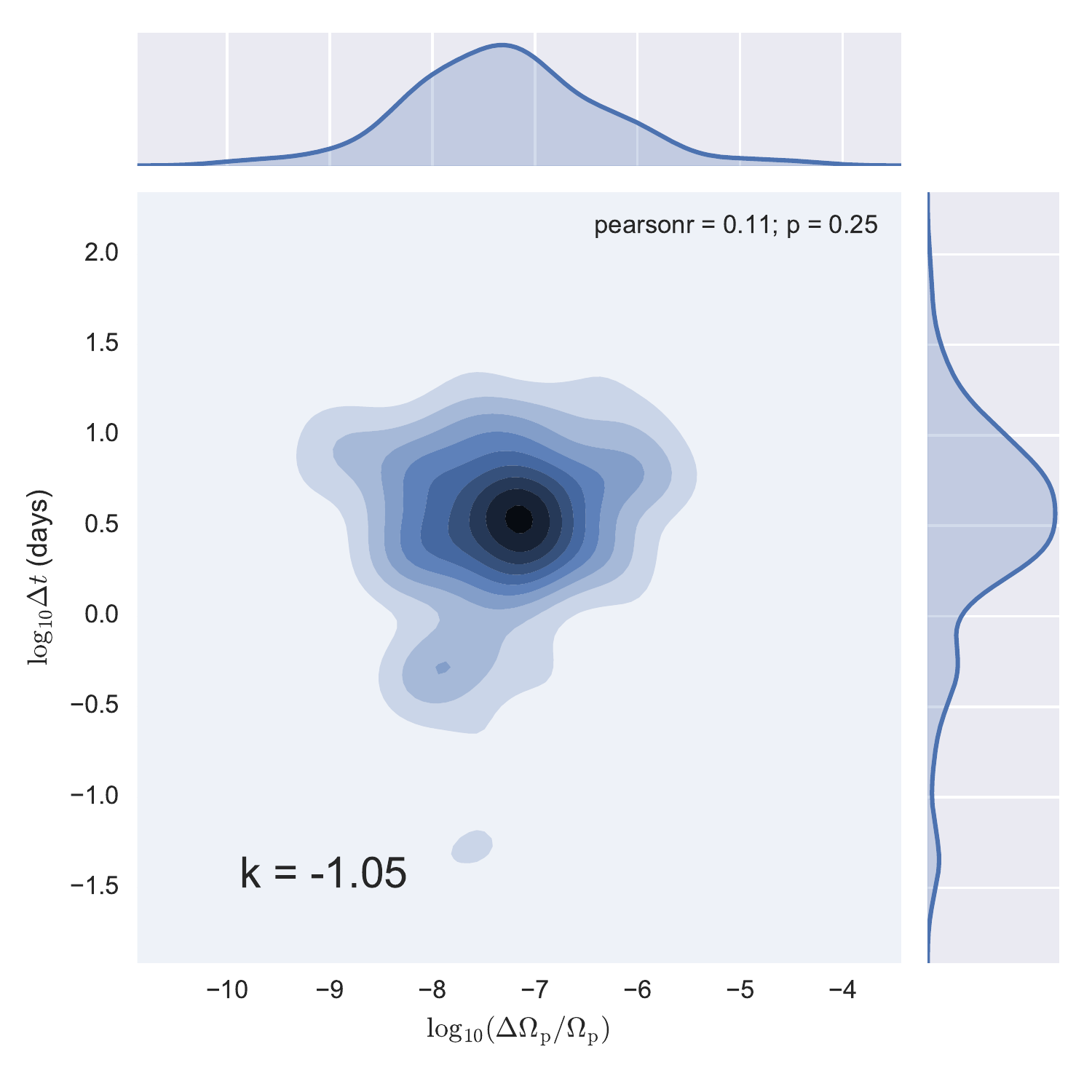}}
\caption{Plots of waiting time ($\Delta t$) versus size ($\Delta\Omega_\mathrm{p}/\Omega_\mathrm{p}$) distributions for microscopic waiting times of $t_w=0.1$ days, both in the case of $k=-1.5$ (left panel) and $k=-1.05$ (right panel). In both cases the distributions differ significantly from power-laws for the sizes and exponentials for the waiting times. There is also no significant correlations between sizes and waiting times, as can be seen from the Pearson $r$ and $p$ coefficients. However, for $k=-1.5$ there is an apparent excess of larger glitches which is associated with a weak quasi-periodicity.}\label{fig4}
\end{figure*}

Figures \ref{fig1} and \ref{fig2} also reveal that a steeper power-law with index $k=-1.5$, rather than $k=-1.05$, on the microphysical level leads to a macroscopic distribution of glitch sizes that appears more narrowly peaked around larger glitches sizes. This is due to the fact that a steeper power law leads to more small events that are not picked up by the glitch finder. Size distributions for $k=-1.5$ thus differ significantly from a power-law and show an abundance of larger glitches. Fitting a power law to the full distribution returns an index $k>-1$, and even cutting off the low end of the distribution returns a fit that is rejected by a Kolmogorov-Smirnov (KS) test. For $k=-1.5$ the waiting time distribution is also skewed towards longer waiting times compared to the $k=-1.05$ case, even if the microphysical choice of waiting times is the same. Furthermore for $k=-1.5$ larger glitches are associated with longer waiting times, as can be seen in figure \ref{fig4}, where we plot the waiting time versus size distribution for a microscopic waiting time of  $t_w=0.1$ days for the two cases $k=-1.5$ and $k=-1.05$. However, we find no significant correlation between waiting times and sizes of glitches, which is consistent with the lack of any such correlation in the observed distributions of glitches in pulsars \citep{Melatos08}, with the sole exception of the pulsar J0537-6910, in which \citep{Mid06} suggest the existence of a correlation between the size of a glitch and the waiting time to the next event. 

\begin{figure}
\centerline{\includegraphics[scale=0.5]{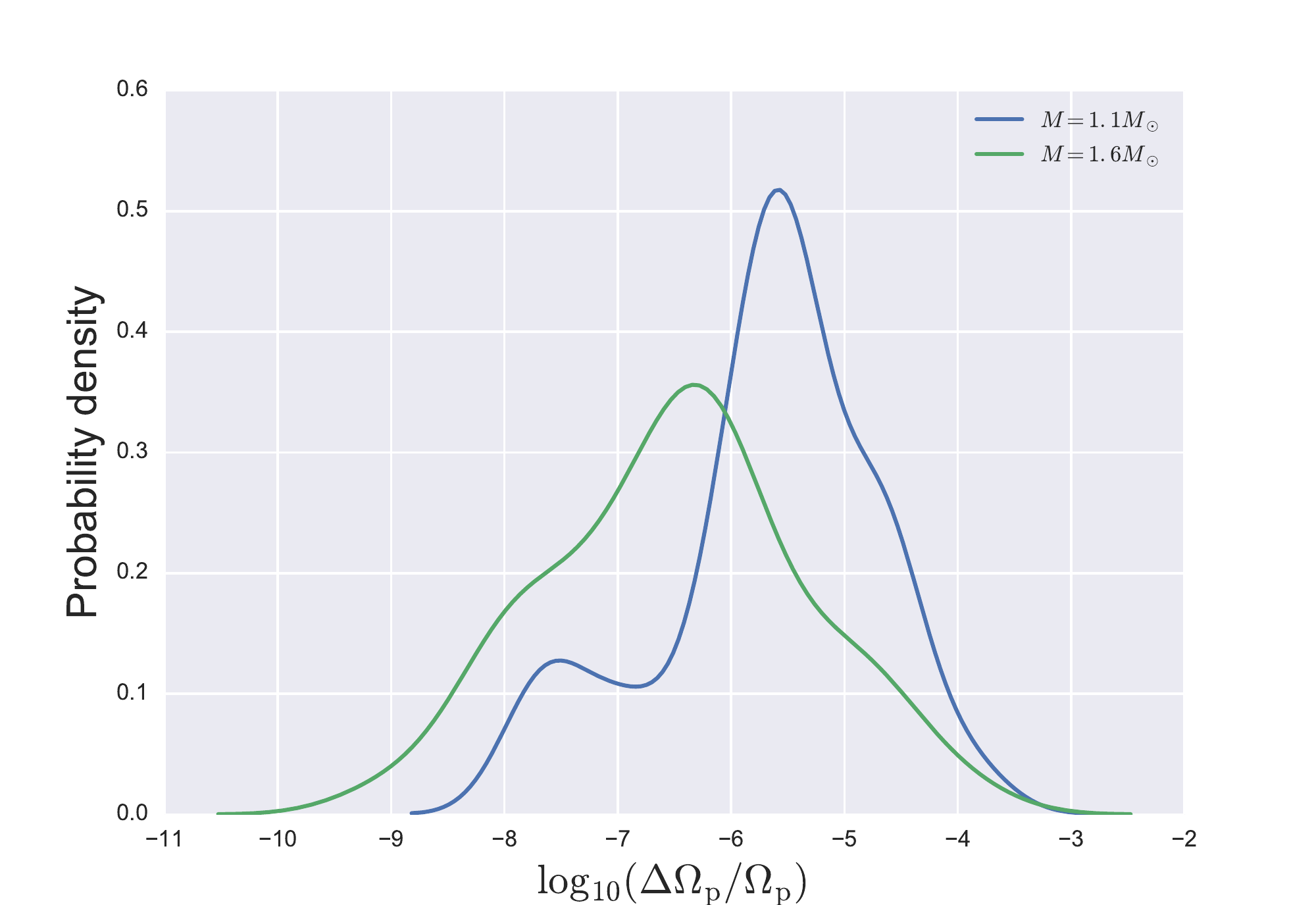}}
\caption{Probability distribution function for glitch sizes $\Delta\Omega_\mathrm{p}/\Omega_\mathrm{p}$ for a model with microscopic power-law index $k=-1.5$ and waiting time $t_w=0.1$ days, for two different neutron star models, one with $M$=$1.1$ M$_\odot$ and $R=14$ km, and the other with $M$=$1.6$ M$_\odot$ and $R=10$ km. The probability distribution function is obtained a gaussian kernel density estimator. We can see that for lower mass stars the distribution is more narrowly peaked around higher values for the glitch size.}\label{fig6}
\end{figure}

Mass is also a key factor in determining glitch sizes, with lower mass stars exhibiting generally larger glitches than high mass stars, as can be seen from figure (\ref{fig6}), where we compare a $M=1.1$ M$_\odot$ and a $M=1.6$ M$_\odot$ neutron star. The waiting time distributions, on the other hand, show no strong dependance on mass.

We have also tested whether the waiting time distribution depends on changes the spin-down rate $\dot{\Omega}_O$ ( or equivalently the height of the maximum of $\Delta_C$). The snowplow model of \citet{pierre} predicts, in fact, that large glitches should occur quasi-periodically as the system builds up the maximum lag $\Delta_C^{max}$, on a timescale $\tau_g=\Delta_C^{max}/\dot{\Omega}_O$. We test this hypothesis by comparing the waiting time distribution for $k=-1.5$ and $t_w=0.1$ days for $\dot{\Omega}_O=-6\times 10^{-10}$ and $\dot{\Omega}_O=-6\times 10^{-11}$. We find no strong evidence for a correlation between the spin-down rate and the glitch waiting time and we do not recover the relation predicted by \citet{pierre} and a KS test returns a 48 $\%$ probability of the two datasets being drawn from the same distribution.  We caution the reader, however, that we have assumed that there is only a single cause for glitches, namely vortex avalanches. The 'snowplow' mechanism suggests that large glitches occur periodically when a vortex sheet forms in the crust where the pinning forces is strongest. If such a mechanism where active together with regularly occurring avalanches one may expect to see an additional periodicity and abundance of large glitches in the distributions. We intend to investigate this hypothesis in future work.

\bibliographystyle{mn2e}
\bibliography{glitchrev}
\bsp

\label{lastpage}

\end{document}